\documentclass[letterpaper, 10 pt, conference]{ieeeconf}  
\IEEEoverridecommandlockouts                              
\usepackage{graphics} % for pdf, bitmapped graphics files
\usepackage{graphicx}
\usepackage{epsfig} % for postscript graphics files
\usepackage{times} % assumes new font selection scheme installed
\usepackage{makecell}
\usepackage{balance}
\usepackage{color}
\usepackage{epstopdf}
\usepackage[utf8]{inputenc}
\usepackage{amsmath}
\usepackage{amssymb}

\usepackage{booktabs}
\usepackage{mathrsfs}
\usepackage{theorem}
\usepackage{algorithm}
\usepackage{algorithmic}
\usepackage{tabularx}
\usepackage{indentfirst}
\usepackage{float}
\usepackage[hidelinks]{hyperref}

\newtheorem{remark}{Remark}

\newcommand{\tr}{\top} % transpose

\newcommand{\imag}{\mathfrak{i}}
\usepackage[style=ieee,backend=bibtex,url=false,doi=false,isbn=false,date=year,citestyle=numeric-comp,maxbibnames=10,minbibnames=10,maxcitenames=10,mincitenames=10]{biblatex}
\bibliography{paper,bibAlex}
\bstctlcite{IEEEexample:BSTcontrol}

\title{\LARGE \bf Approximations for  Optimal Experimental Design \\ in Power System Parameter Estimation
}

\author{Xu Du, Alexander Engelmann, Timm Faulwasser and Boris Houska  % <-this % stops a space
	\thanks{XD and BH are with the School of Information Science and Technology, ShanghaiTech University, Shanghai, China.
		{\tt \{duxu, borish\}@shanghaitech.edu.cn }
	}%
\thanks{XD is also with Shanghai Institute of Microsystem and Information Technology, Chinese Academy of Sciences, as well as the University of Chinese Academy of Sciences, China.}
	\thanks{TF and AE are with the Institute for Energy Systems, Energy Efficiency and Energy Economics, TU Dortmund University, Germany. 
		{\tt \{alexander.engelmann, timm.faulwasser\}@ieee.org,}}%
}

\begin{document}
	
\maketitle
\thispagestyle{empty}
\pagestyle{empty}

%%%%%%%%%%%%%%%%%%%%%%%%%%%%%%%%%%%%%%%%%%%%%%%%%%%%%%%%%%%%%%%%%%%%%%%%%%%%%%%%
\begin{abstract}
	This paper is about computationally tractable methods for power system parameter estimation and Optimal Experiment Design (OED). 
	{The main motivation of OED is  to increase }the accuracy of power system parameter estimates {for a given number of batches}. {One issue in OED, however, is } that solving the  OED problem for larger power grids turns out to be computationally expensive and, in many cases,  computationally intractable. Therefore, the present paper proposes three numerical approximation techniques, which increase the computational tractability of OED for power systems. These approximation techniques are benchmarked on a 5-bus and a 14-bus case studies.

\emph{Keywords:}
Power Systems, Parameter Estimation, Optimal Experiment Design, Admittance Estimation

\end{abstract}

%%%%%%%%%%%%%%%%%%%%%%%%%%%%%%%%%%%%%%%%%%%%%%%%%%%%%%%%%%%%%%%%%%%%%%%%%%%%%%%%
\section{Introduction}
The power system industry is facing a variety of challenges such as supply diversification, reducing carbon emissions, secure network access for renewable energy and electricity market pricing.
%Therefore, the optimal decision and parameter estimation in power grid will greatly affect the actual effect of power grid operation.
Techniques based on online optimization are among the most promising approaches for addressing these challenges.
These techniques range from classical Optimal Power Flow~(OPF) problems \cite{zhu2015optimization}, over optimal reactive power dispatch \cite{frank2012optimal1}, to reactive power planning  \cite{frank2012optimal2}.

In the above approaches, the admittance of the power grid is typically treated as known. 
However, parameters are often unknown in practice and may even vary over time, e.g., due to temperature changes.
Hence, online parameter estimation is used. 
Approaches based on multiple measurements snapshots via Recursive Least Squares~(RLS) has been considered in \cite{Bian2011,Slutsker1996,lateef2019bus,saadeh2016estimation}.
However, RLS-based estimation is sometimes limited in accuracy. 

More recently, Optimal Experimental Design (OED) was applied as an alternative to pure RLS \cite{Du2020,fabbiani2020identification,fabbiani2021identification}. 
Here, the main idea is to choose active/reactive power inputs of the generators such that the  amount of extracted information in the estimation step is maximized. 
{However,} doing so may lead to high costs for system operation, since OED neglects the cost of power generation. 
Having this in mind and inspired by \cite{Houska2015}, \cite{Du2021} proposes an adaptive method for trading-off OED and the OPF cost.
This leads to an excitation of the system, mainly its reactive power, which reduces the cost of optimal estimation substantially compared with pure OED, but still leads to an improved performance compared with RLS.
A drawback of all OED approaches outlined before is their computational intractability for larger grids. 
The objective function in OED requires  inversion of the Fisher Information Matrix (FIM), which does not scale well with the number of buses. 
Moreover, the nonlinear power flow equations lead to additional complexity.

{  In the present paper, we propose approximation techniques to improve the numerical tractability of OED for power system parameter estimation.}
  Specifically, we {develop} an approach based on a Newton-type iteration for inner OED approximation and two outer approximation techniques for approximating the inversion of the Fisher matrix.
  All approximations avoid symbolic matrix inversion in the implementation, which is potentially costly.
  In terms of controlled variables, we focus on the reactive power in order to keep the system operation cost low during the estimation.

 The remaining chapters are structured as follows: \autoref{sec: power system model} reviews  basics of AC power system modeling.
 \autoref{sec:OED} presents the approximation techniques outlined above. \autoref{sec:numerical result} presents numerical case studies for a 5-bus and a 14-bus power grid.
 
  %In \autoref{sec:OED}, we present the approximation techniques outlined above, and we test them on a benchmark example in  \autoref{sec:numerical result}. 
 %the experimental results show that when the precision gap is small, the combination of the Inner Status Approximation and Outer Quadratic Approximated OED can accelerate the operation.

\textit{Notation:}  For $a \in \mathbb{R}^n$ and $\mathcal{C}\subseteq\{1,...,n\}$, ${(a_i)_{i \in \mathcal C}\in\mathbb{R}^{|\mathcal{C}|}}$ collects all components of $a$ whose index $i$ is in $\mathcal C$.  Similarly, for  $A \in \mathbb R^{n \times l}$ and $\mathcal S \subseteq \{ 1, \ldots, n \} \times \{ 1, \ldots, l \}$, $(A_{i,j})_{(i,j) \in \mathcal S}\in\mathbb{R}^{|\mathcal{S}|}$ denotes the {concatenation of} $A_{i,j}$ for all $(i,j)\in\mathcal{S}$. $\imag = \sqrt{-1}$ denotes the imaginary unit, such that  $\mathrm{Re}(z) +\imag\cdot \mathrm{Im}(z)=z \in \mathbb C$, and $\hat a$ denotes the estimated value of $a$. 

\section{AC Power System Model}\label{sec: power system model}
\begin{figure}[t] 
	\centering
	\includegraphics[width=0.7\linewidth]{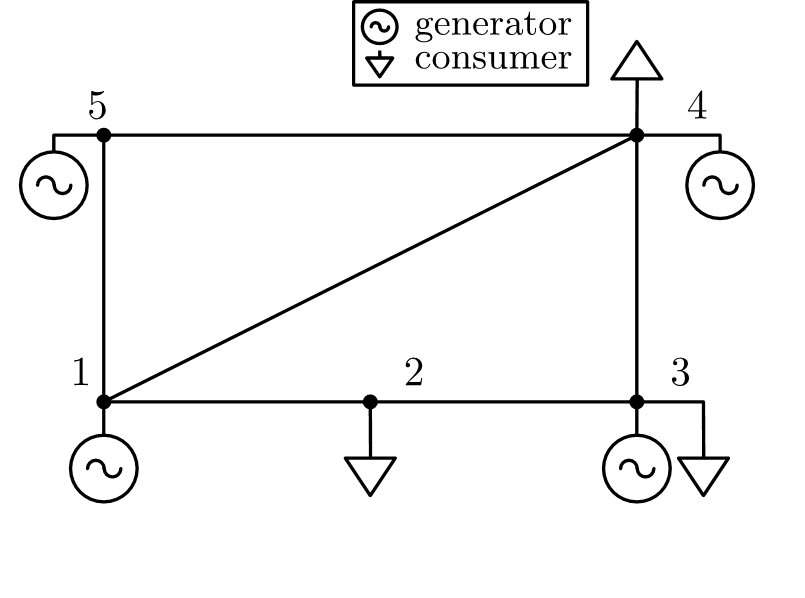}
	\caption{Modiﬁed 5-bus system from \cite{Li2010} with 4 generators and 3 consumers.} 
	\label{fig:ieee5bus}
\end{figure}

Consider a power grid defined by the triple $(\mathcal{N},\mathcal{L},Y)$, where $\mathcal{N} = \{1,2,\dots N\}$ represents the set of buses, $\mathcal{L}\subseteq \mathcal{N}\times \mathcal{N}$ specifies the transmission lines and $Y\in \mathbb{C}^{N\times N}$ denotes the complex admittance matrix
\[
Y_{k,l}\doteq\left\{
\begin{array}{ll}
	\sum\limits_{i \neq k} \left( g_{k,i} + \imag \, b_{k,i} \right) & \text{if} \; k=l, \\  [0.25cm]
	- \left( g_{k,l} + \imag \, b_{k,l} \right) & \text{if} \; k \neq l.
\end{array}
\right.
\]
Here, $g_{k,l}$ and $b_{k,l}$ are the conductances and susceptances of the transmission line $(k,l)\in \mathcal{L}$, which we aim to estimate.
Note that $Y_{k,l}=0$ if $(k,l)\notin \mathcal{L}$.
%In details, the admittance matrix is defined as
%\[
%Y_{i,j}=\left\{
%\begin{array}{ll}
%	\sum\limits_{k \neq i} \left( g_{i,k} + \imag \, b_{i,k} \right) & \text{if} \; i=j, \\  [0.25cm]
%	- \left( g_{i,j} + \imag \, b_{i,j} \right) & \text{if} \; i \neq j,
%\end{array}
%\right.
%\]
%where $g_{i,j}$ and $b_{i,j}$ are the conductances and susceptances of the transmission line $(i,j)\in \mathcal{L}$ respectively. 
The set  $\mathcal G\in \mathcal N$ collects all nodes equipped with generators.  
%In general, not all the buses are connected hence $Y$ matrix can be treated as a \emph{sparse Laplacian matrix}. Figure \ref{fig:ieee5bus} shows the IEEE bus-5 structure and it can be seen that most of the buses are not directly connected. 
Figure \ref{fig:ieee5bus} shows an exemplary 5-bus system with $\mathcal{N}=\{1,\dots,5\}$, and $\mathcal G = \{1,3,4,5\}$.
 %$\mathcal{L}=\{(1,2),(2,3),(3,4),(4,5),(1,5),(1,4)\}$, 

Let $v_k$ denote the voltage amplitude at the $k$-th node and $\theta_{k}$ the corresponding voltage angle. Throughout this paper, we assume that the voltage magnitude and the voltage angle at the first node (the slack node) are fixed, $\theta_1 = \mathrm{const} \quad \text{and} \quad v_1 = \mathrm{const}$. 
This assumption can be made without loss of generality, since the power flow in the network depends on the voltage angle differences $\theta_k - \theta_l$. 
%Similarly, the voltage $v_1$ at the first node is regarded as the reference voltage. 		
Since $\theta_1$ and $v_1$ are given, we define the state of the system as
$
x \doteq \left( v_2, \theta_2, v_3, \theta_3, \ldots, v_N, \theta_N \right)^\tr.
$
Moreover, we have active and reactive power generation of generators $p_k^g$ and $q_k^g$ for all $k\in \mathcal G$.
{The tuple  $d_k = (p_k^d ,\;q_k^d)^\top$ denotes the active and reactive power demand at demand nodes $\mathcal D \subseteq \mathcal N$.}
As an input, we consider only the reactive powers at all but the first generator
$
	%\notag
	u\doteq \left(p_1^\mathrm{g}, (q_k^\mathrm{g})_{k \in \mathcal G }
	\right) \; 
$
{and we assume that the active power generation $\{p_k^g\}_{k \in \mathcal G\setminus \{ 1 \}}$ is fixed.}
Moreover, {{$\label{eq:y}
			y \doteq \left(
				g_{k,l}, % \\[0.16cm]
				b_{k,l} 
			\right)^\top_{(k,l)\in \mathcal{L}} \in \mathbb{R}^{2|\mathcal{L}|} %\notag
$}}
denotes the parameter vector.
Note that {transmission lines}  $(k,l) \notin \mathcal{L}$ are not {considered} in $y$, i.e., {they are not estimated } and thus the sparsity of $Y$ is  considered. The  grid topology is assumed to be known in advance.

The active and reactive power flow over the transmission line $(k,l) \in \mathcal L$ is given by
\begin{align*}
	\Pi_{k,l}(x,y) \doteq v_k^2\hspace{-0.07cm} \left(
		\hspace{-0.25cm}
	\begin{array}{r}
		g_{k,l} \\[0.16cm]
		- b_{k,l} 
	\end{array}
	\hspace{-0.15cm}
	\right) 	\hspace{-0.1cm}-	\hspace{-0.07cm} v_kv_l
		\hspace{-0.13cm}
	\left(
	\hspace{-0.25cm}
	\begin{array}{rr}
		g_{k,l}  & b_{k,l} \\[0.16cm]
		-b_{k,l} & g_{k,l} 
	\end{array}
	\hspace{-0.15cm}
	\right)
	\hspace{-0.15cm}
	\left(
	\hspace{-0.2cm}
	\begin{array}{c}
		\cos(\theta_{k}-\theta_{l}) \\[0.16cm]
		\sin(\theta_{k}-\theta_{l})
	\end{array}
	\hspace{-0.2cm}
	\right)\hspace{-0.1cm}.
\end{align*}
%The relation between the algebraic states, the line parameters and  the transmission line flows for a line $(k,l) \in \mathcal{L}$ is graphically illustrated in \autoref{fig:line}.
The total power outflow  from node $k \in \mathcal{N}$ is given by
\begin{equation}
	\begin{split}
		\sum_{l\in \mathcal N_k} \Pi_{k,l}(x,y) &=
	P_k(x,y) \doteq v_k^2 \sum_{l \in \mathcal N_k}   \left(
	\hspace{-0.25cm}
	\begin{array}{r}
		g_{k,l} \\[0.16cm]
		- b_{k,l} 
	\end{array}
	\hspace{-0.15cm}
	\right)  \notag \\
	& - v_k \sum_{l \in \mathcal N_k} v_l
		\left(
	\hspace{-0.25cm}
	\begin{array}{rr}
		g_{k,l}  & b_{k,l} \\[0.16cm]
		-b_{k,l} & g_{k,l} 
	\end{array}
	\hspace{-0.15cm}
	\right)
	\hspace{-0.15cm}
	\left(
	\hspace{-0.2cm}
	\begin{array}{c}
		\cos(\theta_{k}-\theta_{l}) \\[0.16cm]
		\sin(\theta_{k}-\theta_{l})
	\end{array}
	\hspace{-0.2cm}
	\right), \\
\end{split}
\end{equation}
where $\mathcal N_k \doteq \{ l \in \mathcal N \, \mid \, (k,l) \in \mathcal L \, \}$
denotes the set of neighbors of node $k \in \mathcal{N}$.
%\begin{figure}[t]
%	\centering
%	\includegraphics[width=0.75\linewidth]{figure/line}
%	\caption{Complex voltages and  flow over transmission line  $(k,l) \in \mathcal L$.}
%	\label{fig:line}
%\end{figure}
Thus, the power flow equations can be written in the form 
\begin{equation}
		\label{eq::powerEq}
		\begin{split}
		&P(x,y) = S(u) \; 
	\end{split},
\end{equation}
with {{$\mathrm{dim}(P)  = 2|\mathcal N|$, where}}
$
	P(x,y) \doteq \left ( P_1(x,y)^\tr, \ldots, P_N(x,y)^\tr \right )^\tr,$
and $ S(u) \doteq \left ( S_1(u)^\tr, \ldots, S_N(u)^\tr \right )^\tr.$
Moreover,
\begin{equation*}
	S_k(u)\doteq
				\begin{pmatrix}
			p_k^g\\[0.16cm]
			q_k^g
		\end{pmatrix}-d_k, \; k \in \mathcal D,\;
		S_k(u)\doteq
		\begin{pmatrix}
			p_k^g\\[0.16cm]
			q_k^g
		\end{pmatrix},\; k \notin \mathcal D.
\end{equation*}

\section{Optimal Design of Experiments }\label{sec:OED}

%Next, we recall the basics of optimal experiment design .
%%
%%In this chapter we reviewed the basics of \emph{Fisher Information} and proposed \emph{Inner-Status-Approximation}, \emph{Refined-Linear and Rough-Qudratic Outer Approximation} for avoiding the inverse terms in OED formulation. 
%%
%
%
%\subsection{Maximum-Likelihood Parameter Estimation}
In the following we assume that we can measure  all states~$x$ and the power flow over the transmission lines $\pi_{k,l}$.
%\footnote{\Xu{The choice of measurement function is not unique, as long as the estimator of the parameter is theoretically guaranteed to be unbiased.}} 
Hence, the measurement function is defined by
\[
M(x,y) \doteq \left[ x^\tr, (\Pi_{k,l}(x,y))_{(k,l) \in \mathcal L}^\tr \right]^\tr \;.
\]
We assume additive Gaussian measurement noise with zero mean and given variance $\Sigma \in \mathbb S^{m }_{++}$, i.e.  $\chi \sim \mathcal N(0, \Sigma)$. Hence, the measurements $\eta$ are given by
\begin{align*}
	\eta = M(x,y) + \chi.
\end{align*}
An associated Maximum Likelihood Estimation~(MLE) problem  is given by \cite{Slutsker1996}
\begin{equation}
	\label{eq:MLE}
	\begin{split} 
		\min_{x,y}\;\;& \frac{1}{2}\|M(x,y)-\eta\|_{\Sigma^{-1}}^2+\frac{1}{2}\|y-\hat y\|_{\Sigma_0^{-1}}^2 \\ \quad\mathrm{s.t.}\;\;& 
		%&P_1(x,y) = S_1(u)\\
		P(x,y) = S(u),  \quad \underline x \leq x \leq \overline x .
	\end{split}
\end{equation}
Here, we assume that $\hat y \in \mathbb R^{2|\mathcal L|}$ is a given initial parameter estimate with given variance $\Sigma_0 \in \mathbb{S}_{++}^{2|\mathcal L|}$.

\begin{remark}[Minimal Number of Measurements]	
	Note that the number of measurements we use here, ($4|\mathcal L| + 2|\mathcal N|$), is  not minimal.
	A necessary condition to  determine $y$ uniquely is that there are at least $2|\mathcal L|$ measurements.
	This follows from the  implicit function theorem \cite[Thm 9.28]{Rudin2013}.
	 However, the rank of  $\frac{\partial M}{\partial y}$  also depends on the network topology (e.g. whether there exists islands) and the distribution of measurement devices in the network \cite[Chap. 4]{Abur2004}, \cite{Baldwin1993}. 
{	Approaches for reducing the number of measurements with appropriate measurement placement can be found in \cite{pal2016pmu,manousakis2012taxonomy,donti2019matrix}. }
%	 But observe that reducing the amount of measurements also comes with a worse estimation performance. 
\end{remark}

\subsection{The Fisher Information Matrix and OED}

Next, we derive an approximation for the FIM, which we will use to compute inputs to maximize the information gained in an estimation step.
The FIM characterizes the information content, which can be gained by an experiment. 
It can be expressed as \cite{Houska2015} %\Ae{Is there a rigorous derivation of this formula from the general definition from \url{https://en.wikipedia.org/wiki/Fisher_information} somewhere available?}
\begin{align}
	\label{eq::Fisher}
	\mathcal F(x,y,u) \doteq  \Sigma_0^{-1} + \mathcal{T}(x,y,u)^\tr \Sigma^{-1} \mathcal{T}(x,y,u) \, ,
\end{align}
where
\begin{equation}\label{eq:: Chain rule}
	\mathcal{T}(x,y,u) \doteq \frac{\partial}{\partial y} M(x,y) + \frac{\partial}{\partial x} M(x,y) \frac{\partial}{\partial y} x^*(y,u).
\end{equation} 
Note that all derivatives are evaluated at the true parameter $y$.
As the true parameter $y$ is a priori unknown, we  replace $y$ by our current best guess $\hat y$ in the following, which is common practice in the context of OED \cite{Telen2012}.

The power flow equation \eqref{eq::powerEq} has in general multiple solutions. For example, this equation is invariant under voltage angle shifts.
However, if the sensitivity matrix
$
\frac{\partial}{\partial x} P(x,y)
$
has full rank at an optimal solution $(x,y)$ of~\eqref{eq:MLE},  we can use the implicit function theorem to show that a locally differentiable parametric solution $x^*(y,u)$ of Equation~\eqref{eq::powerEq} exists.\footnote{Conditions under which the matrix $\frac{\partial}{\partial x} P(x,y)$ has full-rank can be found in \cite{Hauswirth2018}, where linear independence constraint qualifications for AC power flow problems are discussed in a more general setting.} Thus, the last term in \eqref{eq:: Chain rule} reads
\begin{equation}\label{OED}
	\frac{\partial}{\partial y} x^*(y,u) \doteq- \left[ \frac{\partial}{\partial x} P(x,y) \right]^{-1} \frac{\partial}{\partial y} P(x,y).
	%\notag
\end{equation}

%
%By taking the derivative of the measure function $M$ with respect to $y$, an auxiliary function $\mathcal T$ can be obtained aims to establish the \emph{Fisher Information Matrix} of the to-be estimated parameters from chain rule \cite{Houska2015},
%\begin{equation}\label{eq:: Chain rule}
%	\mathcal{T}(x,y,u) = \frac{\partial}{\partial y} M(x,y) + \frac{\partial}{\partial x} M(x,y) \frac{\partial}{\partial y} x^\star(y,u)\;
%\end{equation} 

\begin{figure}[h]
	\centering
	\includegraphics[width=.9\linewidth]{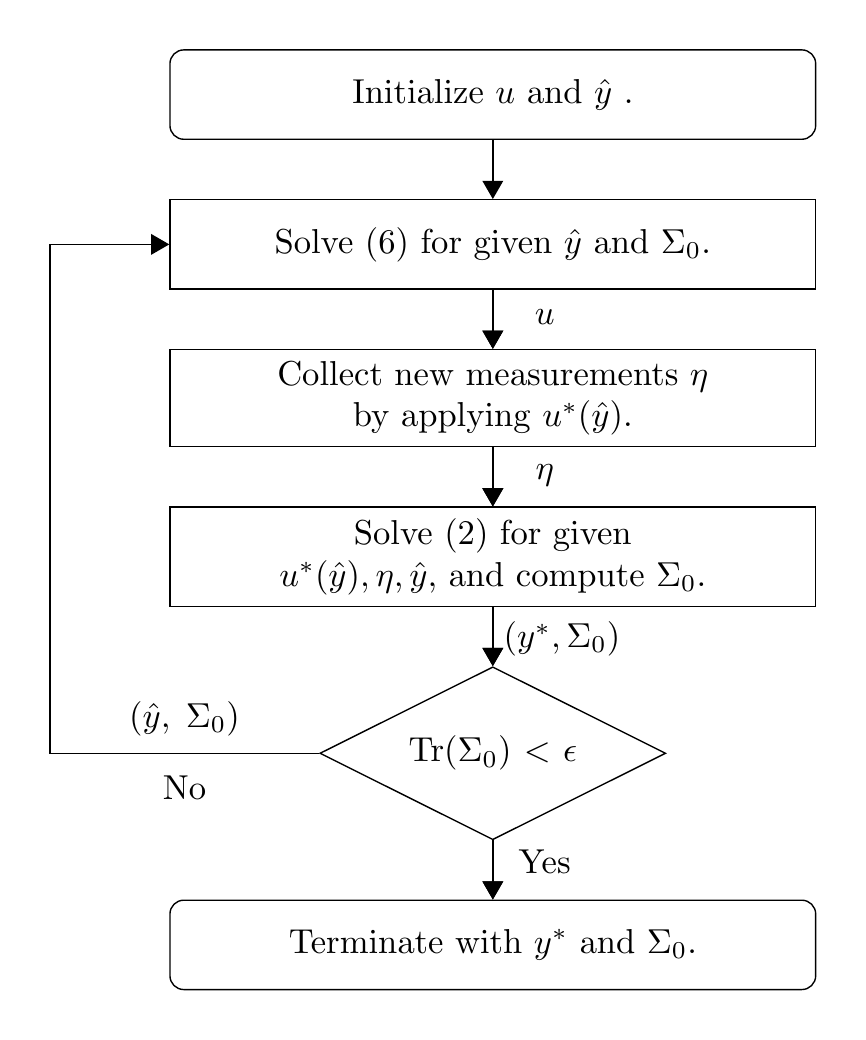}
	\caption{Optimal experiment design for AC power grid admittance estimation.}
	\label{fig:flow}
\end{figure}

%
%In the end, \emph{Fisher information matrix} here is defined as:
%\begin{align}
%	\label{eq::Fisher}
%	\mathcal F(x,y,u) = \mathcal F_0 + \mathcal{T}(x,y,u)^\tr \Sigma^{-1} \mathcal{T}(x,y,u) \, ,
%\end{align}
%whose inverse can be treated as a linear approximation of the variance-covariance matrix of the parameters \cite{kay1993fundamentals}.
Since $\mathcal F$ is a {{mapping}}  $(\mathbb{R}^{n_x},\mathbb{R}^{2|\mathcal L|},\mathbb{R}^{n_u})\rightarrow \mathbb{R}^{2|\mathcal L|\times 2|\mathcal L|}$, we have to choose a scalar criterion for characterizing the information content. 
Typical choices are the $A$-criterion, the $D$-criterion or the $E$-criterion.
For details on their advantages/disadvantages and interpretations we refer to \cite{Telen2012,Telen2013}.
Here, we choose the $A$-criterion, which minimizes the trace of the inverse of the (approximate)  FIM. 
Thus, the OED problem is given by \cite{Du2020}
\begin{equation}
	\label{eq:OED}
	\begin{split} 
	(x^* (\hat y), u^* (\hat y)) 	 = &\arg\min_{x,u}\;\; \mathrm{Tr}([\mathcal F(x,\hat y, u)]^{-1}) \\ \quad\mathrm{s.t.}\;\;&  		\left\{
		\begin{array}{l}
		%	P_1(x,\hat y) = S_1(u)\\[0.16cm]
			P(x,\hat y)=S(u)\\[0.16cm]
%			\underline u_1 \leq P_1(x,\hat y) + \left(
%			\begin{array}{c}
%				p_1^\mathrm{d} \\[0.16cm]
%				q_1^\mathrm{d}\\
%			\end{array}
%			\right) \leq \overline u_1\\
	%\underline p_1^g \leq p_1^g \leq \overline p_1^g\\[0.20cm]
%\underline q_1^g \leq q_1^g \leq \overline q_1^g\\[0.20cm]
			\underline u \leq u \leq \overline u \\[0.16cm]
			 \underline x \leq x \leq \overline x 
			\end{array}\right.\\[0.16cm]
		 	\end{split}
\end{equation}
{The overall OED algorithm is shown in \autoref{fig:flow}.}\footnote{{Note that in contrast to \cite{Du2020}, we omit the regularization term to avoid input chattering. As an alternative one can simply stop the OED algorithm once the desired variance is reached.}}% I delate the flow chart and cite once more our IFAC paper for shorter paragraph

%The prediction of variance of the to-be estimated parameters is influenced by $u$ such that the key idea is by optimizing the trace of the inverse matrix of Fisher information matrix to extract more information, experiments were designed to obtain smaller variance of estimated parameters.
%Note that equation \eqref{eq::powerEq} take care of all the power balance of the power grid such that the consumers no affected by this operation.

%\Ae{\begin{rem}[Exploiting Sparsity in OED]
%	Inhalt...
%\end{rem}}

\section{{Approximation Techniques}}
% \Xu{Different from the title of the next chapter. The purpose of this chapter is to present some basic approximations, and the next chapter is to reconstruct the OED problem.

There are two main computational difficulties, which make solving \eqref{eq:OED}  challenging: a) we have to invert $\partial P/\partial x$ in  \eqref{OED} and, b) we have to compute the inverse of $\mathcal F$.
In the following we will explore approaches for reducing the computational complexity.

\subsection{Approximation via Inner Iterations} \label{sec: Inner Status Approximation}
%\subsubsection*{Approximation of $x^\star(\hat x,y,u)$ via Newton's method } \label{sec: Inner Status Approximation}
The inverse in~\eqref{OED} is difficult to compute in a symbolic or automatic differentiation context---especially if $N$ is large. 
Therefore, we replace the decision variable $x$ with an approximation of $\tilde{x}^*(\hat x, \hat y, \hat u)$ that is defined by an iterative procedure.

%\[\frac{\partial x^*}{\partial p}\approx %\frac{x^*(p+\Delta p)-x^*(p)}{\Delta p} \]
%Newton's method is one of a classical methods for roots searching of power flow equations. Here we use a similar idea by using the current corresponding value Jacobian to hide the nonlinear relation of the power flow equations with respect to $x,\; y$ and $u$ into the approximated status function $x^*(\hat x, y,u)$.
Observe that \eqref{eq::powerEq} is a set of nonlinear equations, which can be solved locally using Newton-type method.
Define $G(x,y,u)=P(x,y)-S(u)$.
Newton-type method requires the evaluation and inversion of  $\frac{\partial G}{\partial x}( x, y, u)$ in each step.
This is expensive, and, hence we use a constant Jacobian approximation at the current iterate $\frac{\partial G}{\partial x}( x, y, u) \approx  \frac{\partial G}{\partial x}( \hat x, \hat y, \hat u)$ instead of an exact Jacobian. 
Algorithm~\ref{alg:Parametric Nonlinear states I} summarizes the resulting algorithm for $K$ iterations.
\begin{algorithm}[tbhp!]
	\caption{Newton-type iteration for  $\tilde{x}^*(\hat x, \hat y,  u)$ }
	\textbf{Input:} Current iterates $(\hat x, \hat y$, $\hat x)$, set $x_0^* = \hat x$.\\
	\textbf{For $k=0,\dots, K$:}
	\begin{itemize}
%		\item[1)] \textit{Newton-type method:}  
		\item[] \[x^*_{k+1}=x^*_k -\left[\frac{\partial G}{\partial x}(\hat x, \hat y, \hat u)\right]^{-1} G(x^*_k, \hat y, u)\]
%		\item[2)] \textit{Update:} 
\vspace{-.4cm}
		\item[] \[k \leftarrow k+1 \] 
	\end{itemize}
	\textbf{Output:} {{$\tilde{x}^*(\hat x,\hat y,u) = x_{K+1}^*$}
	\label{alg:Parametric Nonlinear states I}}
\end{algorithm}
{{Note that from \eqref{eq:: Chain rule}, we obtain an approximation of $\mathcal T,$
}}
\begin{align}
	\mathcal{\tilde T}(\hat x,\hat y,u) \doteq& \frac{\partial}{\partial y} M(\tilde{x}^*(\hat x,\hat y,u),y). \label{eq:tildeT}
\end{align}
Observe that $	\mathcal{\tilde T}$ is different from $\mathcal T$ since $\tilde{x}^*(\hat x,y,u)$ is no longer an independent variable, which  implicitly considers~\eqref{eq::powerEq}.
Thus, we define an approximation of the FIM%from \eqref{eq::Fisher}
\begin{equation}
	\label{eq::Fisher2}
	\mathcal{\tilde{F}}(\hat x,\hat y,u) \doteq \Sigma_0^{-1} + \mathcal{\tilde{T}}(\hat x,\hat y ,u)^\top \Sigma^{-1} \mathcal{\tilde{T}}(\hat x,\hat y ,u).
\end{equation}

\subsection{Fisher linearized  Approximation}

%\Ae{Outer-Refined-Linear Approximated OED}
%\label{sec:Outer-Refined-Linear Approximated OED}
%In the classical OED formulation \eqref{eq:OED}, not only an inverse term of Jacobian matrix of $P$ corresponds with $x$ appears, but also the object function of OED itself is also an inverse expression. Thanks for the corresponding approximated technology in matrix theory, we introduce the approximation of the trace inverse of the \emph{Fisher Information Matrix} as % \Ae{Where is this formula from? Give a reference!)}

Next, we derive an approximation of $ \mathrm{Tr} ( \tilde {\mathcal F}^{-1})$.
%Consider a mapping $\mathbb{R}^{n} \rightarrow \mathbb{R}^{m \times m}, x\mapsto A(x)=(a_{ij})(x)$. Then, define $(\frac{\partial A}{\partial x})_{ijk} \doteq \partial a_{ij}/\partial x_k \in \mathbb{R}^{m\times m\times n}$. Moreover, for a 3-tensor $B=(b_{ijk}) \in \mathbb{R}^{m \times p \times n}$, and a vector $x=(x_k)\in \mathbb R^n$, define $Bx\doteq(\sum_k b_{ijk}x_k)$. With that,
 We have {{\begin{align}
 			\mathrm{Tr}& (  \mathcal{\tilde{F}}(\hat x,\hat y,u ) ^{-1} ) \notag \\
 			=&\mathrm{Tr}(\mathcal{\tilde{F}}(\hat x,\hat y, \hat u)^{-1}) +\mathrm{Tr}\left[\frac{\partial \mathcal{\tilde{F}}^{-1}}{\partial  u}\cdot(u-\hat u)\right]+ O(\|u-\hat u\|^2)\notag \\
 			=&\mathrm{Tr}(\mathcal{\tilde{F}}(\hat x,\hat y, \hat u)^{-1}) + O(\|u-\hat u\|^2)   \notag\\
 			& -\mathrm{Tr}\left(\mathcal{\tilde{F}}(\hat x,\hat y, \hat u)^{-1}\left[\frac{\partial \mathcal{\tilde{F}}}{\partial u}\cdot(u-\hat u)\right]\mathcal{\tilde{F}}(\hat x,\hat y, \hat u)^{-1}\right) \label{eq:objApprox}  \\
 			%	=&\mathrm{Tr}(\mathcal{\tilde{F}}(\hat x,\hat y, \hat u)^{-1}) + O(\|u-\hat u\|^2)\notag\\ &-\mathrm{Tr}(\mathcal{\tilde{F}}(\hat x,\hat y, \hat u)^{-1}[\mathcal{\tilde{F}}(\hat x,\hat y, u)-\mathcal{\tilde{F}}(\hat x,\hat y, \hat u)] \mathcal{\tilde{F}}(\hat x,\hat y, \hat u)^{-1})\notag\\
 			=& 2 \mathrm{Tr}(\hspace{-1pt}\mathcal{\tilde{F}}(\hat x,\hat y, \hat u)^{-1}\hspace{-1pt})\hspace{-2pt}  \notag\\
 			&	-\hspace{-3pt} \mathrm{Tr}(\hspace{-1pt}\mathcal{\tilde{F}}(\hat x,\hat y, \hat u)^{-1}\hspace{-2pt} \mathcal{\tilde{F}}(\hat x,\hat y, u) \mathcal{\tilde{F}}(\hat x,\hat y, \hat u)^{-1})\hspace{-2pt}+\hspace{-2pt}O(\|u-\hat u\|^2) 	, \notag
 		\end{align}
 		where we used a first-order Taylor expansion  and $\mathrm{Tr}(A+B) = \mathrm{Tr}(A) +\mathrm{Tr}(B)$ in the first row,  \cite[ Eq. (59)]{IMM2012-03274} in the second row, and again a Taylor expansion in the third row.
 		Note that $\frac{\partial \mathcal{\tilde{F}}^{-1}}{\partial  u} \in \mathbb R^{2|\mathcal L|\times 2|\mathcal L| \times \mathcal{G}}$ is a tensor with
 		\[\frac{\partial \mathcal{\tilde{F}}^{-1}}{\partial  u}\cdot(u-\hat u)=\mathcal{\tilde{F}}^{-1}\left[\sum_{i}\frac{\partial\mathcal{\tilde{F}} }{\partial u_i}\cdot(u_i-\hat u_i)\right]\mathcal{\tilde{F}}^{-1}.\] }}

%\begin{equation*}
%	\begin{split}
%		&\mathrm{Tr} (  \mathcal{\tilde{F}}(\hat x,\hat y,u ) ^{-1} )\\
%		\approx&\mathrm{Tr}(\mathcal{\tilde{F}}(\hat x,\hat y, \hat u)^{-1})\\ -&\mathrm{Tr}(\mathcal{\tilde{F}}(\hat x,\hat y, \hat u)^{-1}[\frac{\partial \tilde{F}}{\partial u}(u-\hat u)]\mathcal{\tilde{F}}(\hat x,\hat y, \hat u)^{-1})+ O(\|u-\hat u\|^2) \\
%		\approx&\mathrm{Tr}(\mathcal{\tilde{F}}(\hat x,\hat y, \hat u)^{-1})\\ -&\mathrm{Tr}(\mathcal{\tilde{F}}(\hat x,\hat y, \hat u)^{-1}[\mathcal{\tilde{F}}(\hat x,\hat y, u)-\mathcal{\tilde{F}}(\hat x,\hat y, \hat u)]\mathcal{\tilde{F}}(\hat x,\hat y, \hat u)^{-1})\\
%		+&  \hspace{2pt} O(\|u-\hat u\|^2) \\
%		=& 2 \mathrm{Tr}(\hspace{-1pt}\mathcal{\tilde{F}}(\hat x,\hat y, \hat u)^{-1}\hspace{-1pt})\hspace{-2pt}  -\hspace{-2pt} \mathrm{Tr}(\hspace{-1pt}\mathcal{\tilde{F}}(\hat x,\hat y, \hat u)^{-1}\hspace{-2pt} \mathcal{\tilde{F}}(\hat x,\hat y, u) \mathcal{\tilde{F}}(\hat x,\hat y, \hat u)^{-1}\hspace{-2pt})
%	\end{split}
%\end{equation*}

%\Ae{Add a remark that pure OED provides a lower bound on the variance possible by other estimation methods.}% I put it above under the flow chart.

\begin{remark}
	Notice that \eqref{eq:objApprox} can be interpreted as a weighted \emph{T-criterion} \cite[Chapter 6.5]{pukelsheim2006optimal}.%, and the latter can be regarded as maximizing the trace of FIM directly. We use \emph{A-criterion} instead of \emph{T-criterion} for two reasons: a) \emph{T-criterion} has no clear physical meaning but \emph{A-criterion} represents the sum of the variances of the parameters, b) \emph{T-criterion} has unstable numerical performance in practice.
\end{remark}

\section{{{OED Reformulations}}}
In this section we use the approximations from  the previous section to reformulate the OED problems. Three options for doing so are discussed.

\subsection{Fisher Linearized  Approximation OED}\label{sec:Outer-Refined-Linear Approximated OED}
Using both approximations from the previous section, we  rewrite \eqref{eq:OED} as
%\Ae{This is strange... isn't $q^g_k$ included in u?! Why do we need an extra constraint? Isn't $p^g$ fixed? Why do we need a constraint in this case?!} %I fixed equation (6) (10) (11) and (12)
\begin{align}
\label{eq::OED}
		\min_{u}&- \mathrm{Tr}(\mathcal{\tilde{F}}(\hat x, \hat y, \hat u)^{-1} \mathcal{\tilde{F}}( \hat x, \hat y, u) \mathcal{\tilde{F}}(\hat x, \hat y, \hat u)^{-1}) \notag\\
		\text{s.t.}&   		\left\{
		\begin{aligned}
%				\underline u_1 \leq P_1(x^*(\hat x,\hat y,u),\hat y) + \left(
%		\begin{array}{c}
%			p_1^\mathrm{d} \\[0.16cm]
%			q_1^\mathrm{d}\\
%		\end{array}
%		\right) \leq \overline u_1\\[0.16cm]
	%\underline p_1^g &\leq p_1^g \leq \overline p_1^g,\quad 
%	\underline q_1^g &\leq q_1^g \leq \overline q_1^g\\[0.20cm]
		\underline u &\leq u \leq \overline u,\\  
		\underline x &\leq \tilde x^*(\hat x,\hat y,u) \leq \overline x .
	\end{aligned}\right. 
\end{align}
Observe that we only have $u$ as decision variables, which reduces the problem dimension.
We neglect the term $2 \mathrm{Tr}(\hspace{-1pt}\mathcal{\tilde{F}}(\hat x,\hat y, \hat u)^{-1}$ in \eqref{eq:objApprox} since a constant offset in the objective does not change the minimizer.
Note that the variables $(\hat x,\hat y, \hat u)$ are fixed to their current iterates in the above problem, and $\tilde{x}^*(\hat x,\hat y,u)$ is given by Algorithm~\ref{alg:Parametric Nonlinear states I}.

%Observe that the above problem formulation avoids the inversion of $\tilde F$ and also avoids nonlinear equality constraints.
The essence of the above method is to hide the power flow equations via Algorithm~\ref{alg:Parametric Nonlinear states I}. As the number of  iterations $K$ increases, the  complexity of the objective function of~\eqref{eq::OED} increases, although the external form is concise. In the next subsection, we use a different approach, which is less accurate but also less costly.

\subsection{Quadratic Approximation OED}
%\Ae{Please use a less complicated name here. }
%\subsubsection*{Linearized version}
\label{sec:Outer-Rough-Quadratic Approximated OED}

%\Ae{
%	\emph{Outer-Refined-Linear Approximated OED}}

Observe that the objective in \eqref{eq::OED}  is nonlinear. % with higher-order terms.
%As an alternative, we provide a quadratically approximated version 
A quadratic approximation is given by
\begin{align}
	\label{eq::approximated OED}
		\min_{u}&\;\; \frac{1}{2}u^\top H(\hat{x}, \hat{y}, \hat u) u + J(\hat{x}, \hat{y}, \hat u)u \notag\\\vspace{2cm}
		\text{s.t.}& \left\{
		\begin{array}{l}
%		\underline u_1 \leq P_1(x,\hat y) + \left(
%		\begin{array}{c}
%			p_1^\mathrm{d} \\[0.16cm]
%			q_1^\mathrm{d}\\
%		\end{array}
%		\right) \leq \overline u_1\\[0.16cm]
%\underline p_1^g \leq p_1^g \leq \overline p_1^g\\[0.20cm]
%\underline q_1^g \leq q_1^g \leq \overline q_1^g\\[0.20cm]
		\underline u \leq u \leq \overline u,\\[0.16cm]
		\underline x \leq C(\hat{x}, \hat{y}, \hat u)u+\hat{x} \leq \overline x\\
		\end{array}\right.
\end{align}
where
\begin{align*}
		J(\hat{x}, \hat{y},\hat u)\doteq&\frac{\partial E}{\partial u}(\hat{x}, \hat{y}, \hat u), \quad 
		H(\hat{x}, \hat{y}, \hat u)\doteq\frac{\partial^2 E}{\partial u^2}(\hat{x}, \hat{y}, \hat u)\\
			C(\hat{x}, \hat{y}, \hat u)\doteq&\frac{\partial \tilde x^*(\hat x,\hat y,u)}{\partial u}(\hat{x}, \hat{y}, \hat u)
\end{align*}
and $E(\cdot)$ represents the objective function from \eqref{eq::OED}.
Note that the above problem is a Quadratic Program (QP), which can be solved efficiently by standard QP solvers.%which is potentially fast in computation.

%here $\hat u^*(\hat y)$ denotes the optimal input from the last iteration, while the \emph{Quadratic Approximated OED} is represented as

%\Ae{What is $u^\star$? Is it defined somewhere? What happens to the power flow equations?! Shouldn't they also be linearized?}

%Except the first symbolic evaluation of $H$ and $J$, \emph{Quadratic Approximated OED} has almost the same computation complexity as \emph{DC Optimal Power Flow} which is only a \emph{QP} problem.% different quasi-Newtonian Hessian approximations such as BFGS may also be used to speed up operations. On the other hand, the system input with moderate change by the new method.

\subsection{Inner Linearized Approximation OED}\label{sec:Inner Linearization Based Approximated OED}
%\Ae{Please use a less complicated name here. }
%In this section, we proposed a relatively different approximation approach compare with the other two.

Computing $x^*(\hat x,\hat y,u)$ via Algorithm~\ref{alg:Parametric Nonlinear states I} can lead to large memory requirements in the context of automatic differentiation, since the expression graph grows with the number of the Newton-type iterations $K$.

As an alternative, we use the nonlinear constraints from~\eqref{eq:OED} combined with the objective approximation  from \eqref{eq:objApprox}.
This leads to
	\begin{align}  \label{eq:OED3}
		\min_{x,u}\;\;& - \mathrm{Tr}(\mathcal{F}(\hat x, \hat y, \hat u)^{-1} \mathcal{F}( x, \hat y, u) \mathcal{F}(\hat x, \hat y, \hat u)^{-1})  \notag \\ \quad\mathrm{s.t.}\;\;&  		\left\{
		\begin{array}{l}
		%	P_1(x,\hat y) = S_1(u)\\[0.16cm]
			P(x,\hat y)=S(u)\\[0.16cm]
%			\underline u_1 \leq P_1(x,\hat y) + \left(
%			\begin{array}{c}
%				p_1^\mathrm{d} \\[0.16cm]
%				q_1^\mathrm{d}\\
%			\end{array}
%			\right) \leq \overline u_1\\
%	\underline p_1^g \leq p_1^g \leq \overline p_1^g\\[0.20cm]
%\underline q_1^g \leq q_1^g \leq \overline q_1^g\\[0.20cm]
			\underline u \leq u \leq \overline u ,\qquad \underline x \leq x \leq \overline x.\\[0.16cm]			 
		\end{array}\right.
	\end{align}
Here,  we use \eqref{eq::Fisher} for evaluating the objective, where we substitute 
 $\frac{\partial}{\partial y} \bar{x}^*( y, u) =- \left[ \frac{\partial}{\partial x} P(\hat x,\hat y) \right]^{-1} \frac{\partial}{\partial y} P(x,y)$ in  \eqref{eq:: Chain rule} for a fixed Jacobian  evaluated at  $(\hat x, \hat y)$.

\begin{remark}[Difference between $\tilde {\mathcal F}$ and $\mathcal F$]
	Notice the difference between $\tilde {\mathcal F}$ and $\mathcal F$: Whereas $\tilde {\mathcal F}$ uses the approximation $\tilde {\mathcal T}$ from \eqref{eq::Fisher2} including the Newton-type iteration from Algorithm~\ref{alg:Parametric Nonlinear states I},  $\mathcal F$  in \eqref{eq:OED} and \eqref{eq:OED3} uses  \eqref{eq::Fisher}, and \eqref{eq:OED3} with  the approximation  $\frac{\partial}{\partial y} \bar{x}^*( y, u) =- \left[ \frac{\partial}{\partial x} P(\hat x,\hat y) \right]^{-1} \frac{\partial}{\partial y} P(x,y)$ in~\eqref{eq:: Chain rule}.
%	By using \emph{approximation via inner iterations}, \eqref{eq::OED} has notation $\mathcal{\tilde{F}}$ in the object function, while $\mathcal F$ is used both in \eqref{eq:OED} and \eqref{eq:OED3} for the same structure inside except the lagging constant Jacobian inverse matrix. The latter approximation is designed to reduce the nonlinearity of $\mathcal F$ and thus speed up OED operations.
\end{remark}

%By using all the technologies introduced in this section, the procedure of \emph{Approximated Optimal Experimental Design} for power grid admittance estimation is shown as Algorithm \eqref{alg:Approximated OED} and Figure \eqref{fig:flow}:

%\newpage
\section{Numerical Results}\label{sec:numerical result}

We illustrate the numerical performance of the  approximations on a modified 5-bus power system from \cite{Li2010} (cf. \autoref{fig:ieee5bus}) and on a 14-bus power system \cite{christie2000power}.

\subsection{Implementation and Data}

The problem data is obtained from the \texttt{MATPOWER} dataset \cite{zimmerman2010matpower}  ignoring  shunt elements. The implementation of OED relies on \texttt{Casadi-v.3.4.5} with \texttt{IPOPT} \cite{Andersson2019} and \texttt{MATLAB 2020b}.
We use Gaussian measurement noise with zero mean and a variance of $10^{-4}$. % which connected the weighted matrix $\Sigma^{-1}$ in equation \eqref{eq:MLE} and 
The initial value of $\Sigma_0^{-1}$ is set to $10^{-16} I$. %All the setting are the same as \cite{Du2021,Du2020}.
We benchmark our method also against MLE with a  standard Gaussian random input for each generator, scaled such that the total reactive power demand is met.  
{For the 14-bus system, the input variance is $0.01\,\mathrm{(p.u.)}^2$.}

% {\color{blue}{Since the OED can cause input changes to the system, the standard RLS keeps the system input constant, which seems unfair. Here we introduce a Gaussian random input for comparison, with a mean of 0 and a variance of 1.}}
%\Ae{Comment on how we generated the random input and why we did that here. Also give the noise level, i.e. standard deviation and the distribution (Gaussian or uniform?).}

\subsection{Numerical Comparison}

Next,  we compare all OED variants numerically: a) classical OED  from \cite{Du2020} using~\eqref{eq:OED}, b) \emph{Fisher Linearized  Approximation OED} (FLA-OED) from \autoref{sec:Outer-Refined-Linear Approximated OED}, c) \emph{Quadratic Approximation OED} (QA-OED) method from \autoref{sec:Outer-Rough-Quadratic Approximated OED}, and, d) \emph{Inner Linearized Approximation OED} (ILA-OED), from \autoref{sec:Inner Linearization Based Approximated OED}.

%For brevity, methods \uppercase\expandafter{\romannumeral1}$\sim$\uppercase\expandafter{\romannumeral4} stand for our four OED variants in the following.
%Method \uppercase\expandafter{\romannumeral1} stands for the classical OED  from \cite{Du2020} using~\eqref{eq:OED}. 
%Method \uppercase\expandafter{\romannumeral2} corresponds to \Ae{\emph{Fisher Linearized  Approximation OED}} (FLA-OED) from \autoref{sec:Outer-Refined-Linear Approximated OED}. 
%Method \uppercase\expandafter{\romannumeral3} refers to the \Ae{\emph{Quadratic Approximation OED}} (QA-OED) method from \autoref{sec:Outer-Rough-Quadratic Approximated OED}, and  method \uppercase\expandafter{\romannumeral4} corresponds to \Ae{ \emph{Inner Linearized Approximation OED}} (ILA-OED), from \autoref{sec:Inner Linearization Based Approximated OED}.  

%\Ae{@Xu: Think of a better naming for the methods and introduce abbreviation in the corresponding sections. Then use the abbreviations also in the plots and the Table. That is much easier to remember than the numbers for the reader. }

\begin{table}[tbhp!] 
%	\centering
	\renewcommand{\arraystretch}{1.3}		
	\caption{Computation time for 200 Batches} 
	\begin{center}
	\begin{tabular}{r||p{1cm}p{1cm}p{1cm}p{1cm}} 
		\hline
		\makecell{OED Method}&\makecell{OED}   & \makecell{FLA}  & \makecell{QA} &\makecell{ILA}\\
		\hline
		\makecell[c]{Time} $[\mathrm{sec}]$ & 19.7926 &36.1696&3.9811 &8.6328\\ 
		\makecell[c]{Sum Variance} $[\mathrm{S^2}]$ & 0.0582 &0.0281& 0.0358 &0.0583\\
		\hline
	\end{tabular}
	\label{table:result1}
	\end{center}
\end{table}
Table \ref{table:result1} shows the computation times with $200$ batches for the 5-bus system.
We use $K=20$  Newton-type iterations in Algorithm~\ref{alg:Parametric Nonlinear states I} for FLA-OED and QA-OED. 
Compared with the classical OED, FLA-OED takes a longer computation time for the same number of batches, but it also achieves a higher accuracy.
\begin{table}[tbhp!] 
	\centering
	\renewcommand{\arraystretch}{1.3}		
	\caption{Computation time for a target sum variance of $0.1$.} 
	\begin{tabular}{r||p{0.7cm}p{0.7cm}p{0.7cm}p{0.7cm}} 
		\hline
		\makecell{Method}&\makecell{OED}   & \makecell{FLA}  & \makecell{QA} &\makecell{ILA}\\
		\hline
		Time $[\mathrm{sec}]$ & 11.1998 &7.7718&1.7076& 4.9239\\ 
		Sum Variance $[\mathrm{S^2}]$ & 0.0992 &0.0996& 0.0983&  0.0990\\
		\hline
	\end{tabular}
	\label{table:target}
\end{table}
Table \ref{table:target} shows the computation time for a given  target sum variance of $0.1$.
Here, one can observe the trade-off between solution accuracy and computation time: the  accurate classical OED method and ILA-OED are faster here, since they require a smaller number of batches to get to the desired variance. 
QA-OED still has the shortest operation time for the given target sum variance. 
QA-OED solves the above problem via quadratic approximation, which leads to  fast computation. 
ILA-OED shows a similar accuracy as the standard OED with less computation time. 
\begin{figure}[tbhp!]
	\centering
	\includegraphics[width=0.9\linewidth]{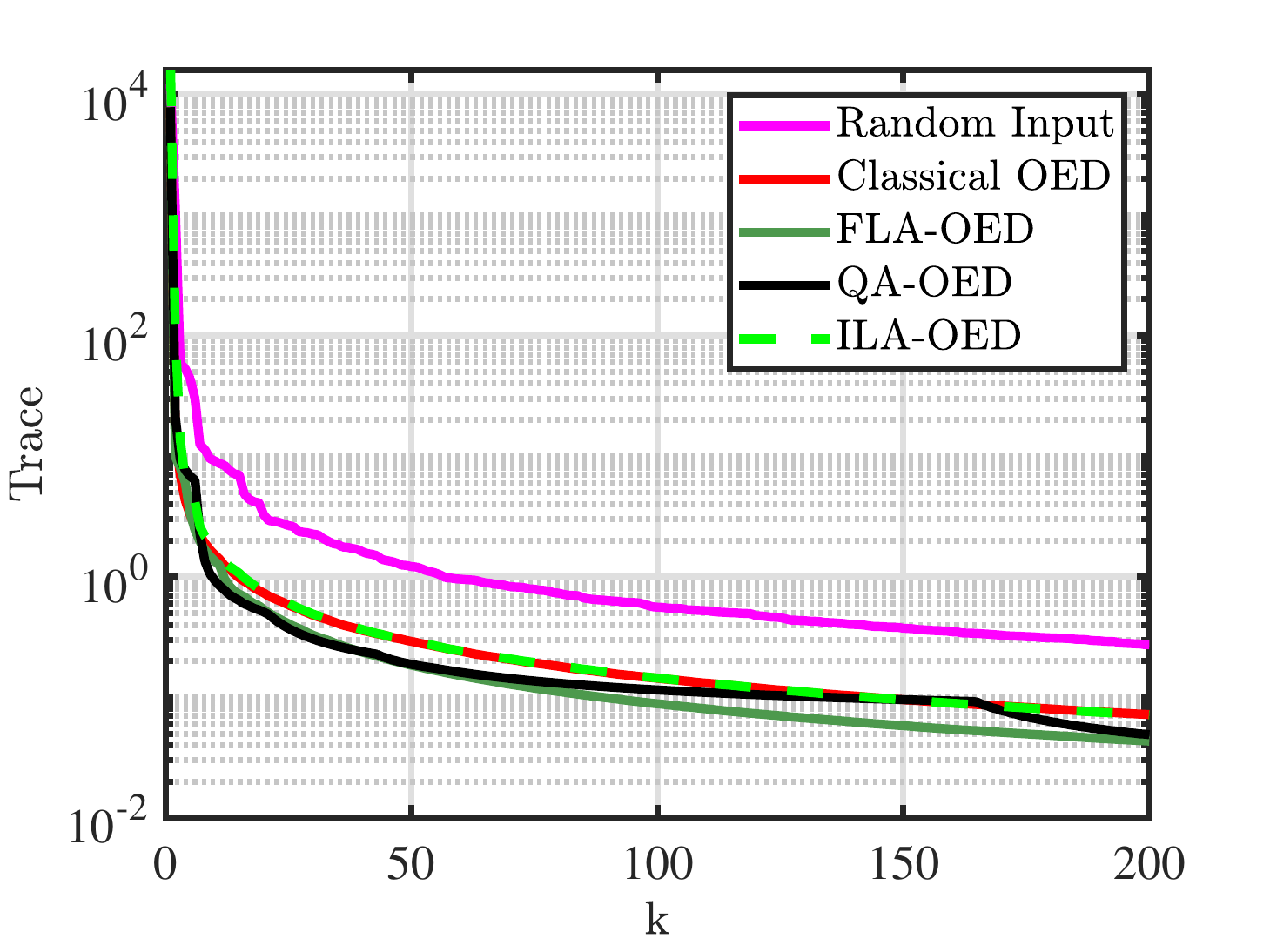}\vspace{-0.4cm}
	\caption{Expected total variance comparison random Input, classical OED, 
		FLA-OED, QA-OED  as well as
		ILA-OED method for the 5-bus power system. }
	\label{fig:variance}
\end{figure}
 \autoref{fig:variance} shows the expected total variance $\mathrm{Tr}(\mathcal{F}(\hat x, \hat u, \hat y)^{-1})$,  where $(\hat x, \hat u, \hat y)$ are fixed to the current iterates.
All four methods lead to similar levels of accuracy and they are more accurate than the pure standard RLS with {random generator input}.

%{\Xu{\begin{remark}
%			Pure OED provides a lower bound on the variance possible by %other estimation methods.
%\end{remark}}}

 %\uppercase\expandafter{\romannumeral1} tooks longer time and a bit higher accuracy than \uppercase\expandafter{\romannumeral2} for more controller input, \uppercase\expandafter{\romannumeral3} takes the longest time for the high complexity implicit expression  of status function. \uppercase\expandafter{\romannumeral4}  solves the complex structural problems brought by \uppercase\expandafter{\romannumeral3} but with a little bit lower accuracy which is still acceptable. Figure \eqref{fig:variance} shows from the evaluation of variance, four methods has almost the same level and far more better than the pure standard RLS.

\begin{figure}[tbhp!]
	\centering
	\includegraphics[width=0.9\linewidth]{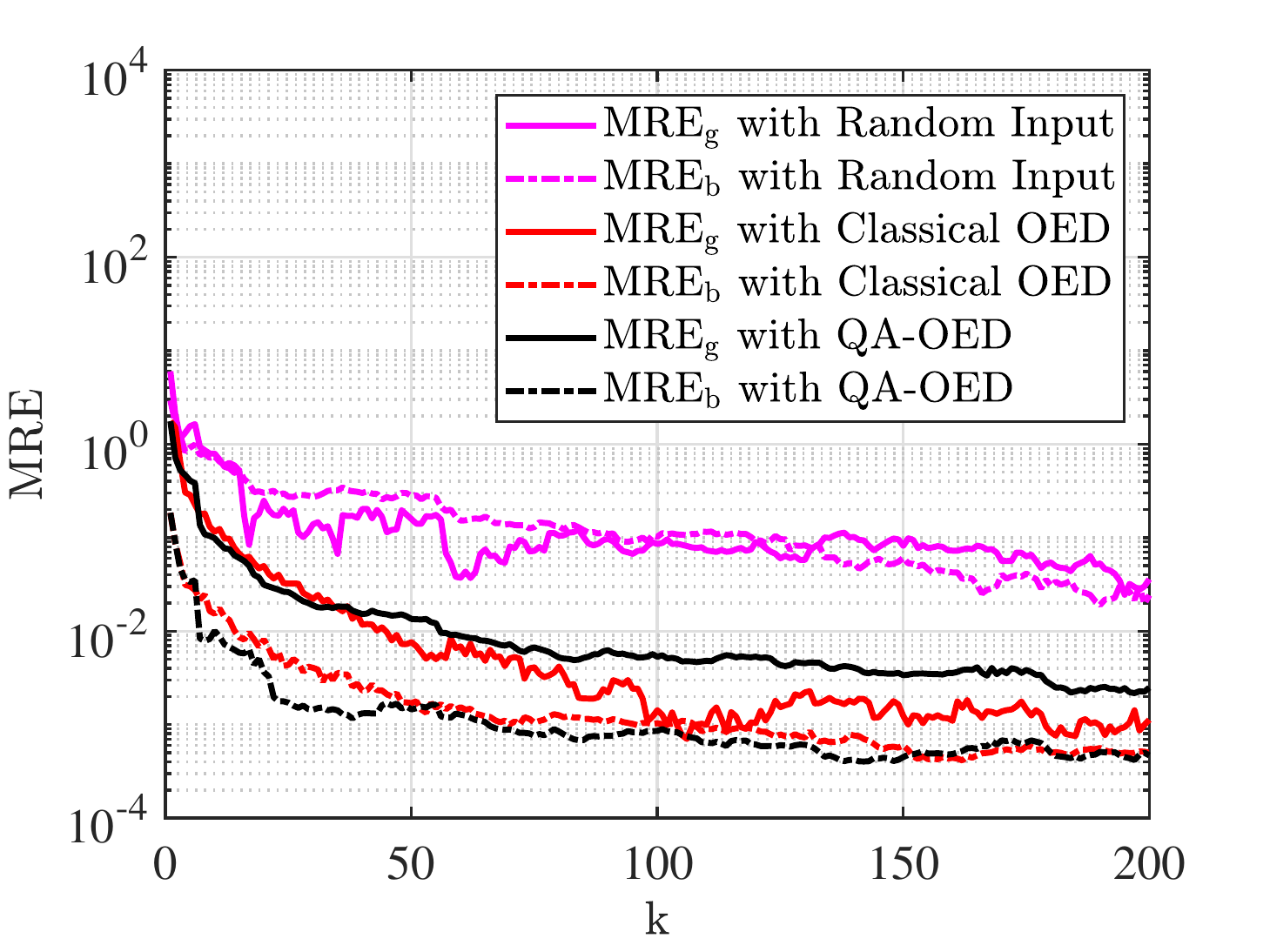} \vspace{-0.3cm}
	\caption{Mean relative errors of $\mathrm{MRE_g}$ (solid line) and $\mathrm{MRE_b}$ (dashed line) for the 5-bus power system. }
	\label{fig:ERROE}
\end{figure}
%\begin{figure}[tbhp!]
%	\centering
%	\includegraphics[width=0.9\linewidth]{figure/MRE_qua.eps} \vspace{-0.3cm}
%	\caption{Mean relative errors of $\mathrm{MRE_g}$ (solid line) and $\mathrm{MRE_b}$ (dashed line) for the 5-bus power system. }
%	\label{fig:ERROE}
%\end{figure}
\autoref{fig:ERROE}  compares the mean relative error (MRE)
\begin{align*}
	\mathrm{MRE}_\mathrm{g}=\frac{1}{|\mathcal{L}|}\sum_{(k,l)\in \mathcal{L}}\frac{|g_{k,l}-\bar g_{k,l}|}{|\bar g_{k,l}|} \;,\\
	\mathrm{MRE}_\mathrm{b}=\frac{1}{|\mathcal{L}|}\sum_{(k,l)\in \mathcal{L}}\frac{|b_{k,l}-\bar b_{k,l}|}{|\bar b_{k,l}|}\;,
\end{align*} 
for {{RLS with random input,}} for classical OED and QA-OED. 
%The blue line represents standard RLS, the red lines represents method \uppercase\expandafter{\romannumeral1}, and the black lines denotes method \uppercase\expandafter{\romannumeral3}. 
The results of FLA-OED and ILA-OED are similar to the ones shown in \autoref{fig:ERROE} and are thus omitted.
%We remark that for  standard RLS, the trend in the MRE (\autoref{fig:ERROE})  is different from the trend for the total variance (\autoref{fig:variance}).
%An explanation for that is that the total variance expresses an expected variance, whereas $\mathrm{MRE}$ is computed based on the knowledge of the ground truth $(\bar g_{k,l},\bar b_{k,l}), (k,l) \in \mathcal L$.
Note that the value of $\mathcal F$ in \eqref{eq::Fisher} increases in each iteration because of the second term, which leads the decrease of its inverse. We refer to \cite[Chapter 7]{Ljung1999} and \cite{Telen2013} for further discussion.

\begin{figure}[tbhp!]
	\centering
	\includegraphics[width=0.9\linewidth,trim={.0cm 1.6cm 0cm 1.3cm},clip]{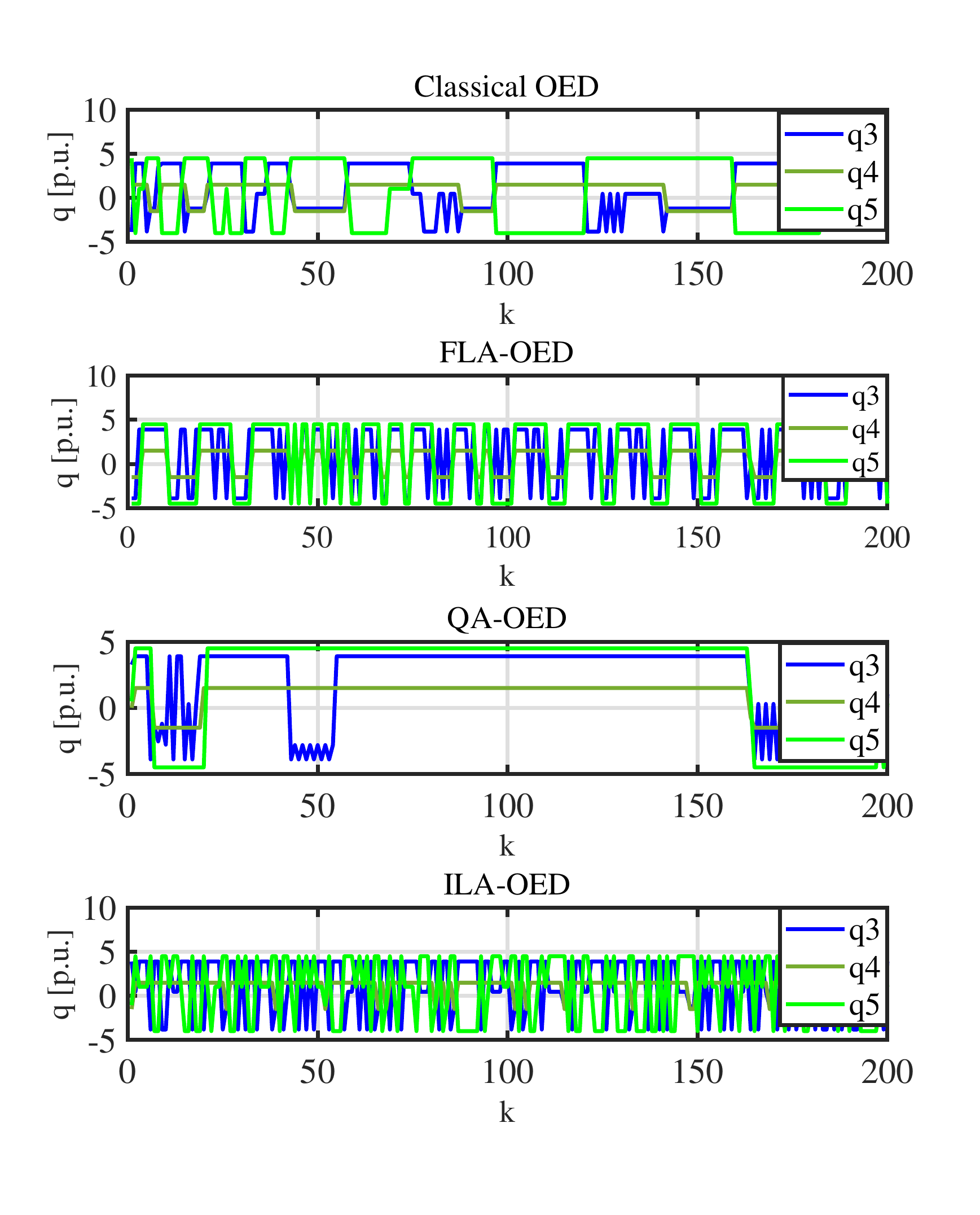}
	\caption{Optimal reactive power inputs for all generators as
		obtained by proposed methods the 5-bus power system.}
	\label{fig:qg_comparison}
\end{figure}

%\begin{figure}[tbhp!]
%	\centering
%	\includegraphics[width=1\linewidth]{figure/q_explict.eps}
%	\caption{Optimal reactive power inputs for all generators as
%		obtained by explict OED.}
%	\label{fig:qg}
%\end{figure}
%
%\begin{figure}[tbhp!]
%	\centering
%	\includegraphics[width=1\linewidth]{figure/q_qua.eps}
%	\caption{Optimal reactive power inputs for all generators as
%		obtained by Algorithm \eqref{alg:Approximated OED}.}
%	\label{fig:qg_qua}
%\end{figure}
\autoref{fig:qg_comparison} shows the  optimal reactive power inputs for all methods.
Here one can see that the reactive power update is less frequent in QA-OED compared with the other three methods, which is benefitial in grid operation.
%\Ae{Note that we update the reactive power here for avoiding mechanical stress in generators and also due to cost reasons, cf. \cite{Du2021} .}

% another good property of method  \uppercase\expandafter{\romannumeral3}: the reactive power update is much smoother than the other three methods which will benefit the actual grid operation.

Table \ref{table:loss of optimality} shows the relative loss of optimality $\Delta(u(\cdot))$ of the first iteration with different methods based on the optimal objective $T^\star$ of classical OED,
\begin{align*}
\Delta(u(\cdot))= \frac{\mathrm{Tr}(\mathcal{F}^{-1}(x(u(\cdot),\hat y)), \hat y, u(\cdot))-T^\star}{T^\star}.
\end{align*}
\begin{table}[tbhp!] 
	\centering
	\renewcommand{\arraystretch}{1.3}		
	\caption{Relative loss of optimality for the first {iteration}} 
	\begin{tabular}{r||p{1.2cm}p{0.7cm}p{0.7cm}p{0.7cm}} 
		\hline
		\makecell{Method}&\makecell{Random}   & \makecell{FLA}  & \makecell{QA} &\makecell{ILA}\\
		\hline
		$\Delta(u(\cdot))$ &\makecell{8.45} &\makecell{0.36}& \makecell{0.36}&  \makecell{4.94}\\
		\hline
	\end{tabular}
	\label{table:loss of optimality}
\end{table}
Here, $u(\cdot)$ denotes inputs for different OED variants.
The relative loss of optimality for the random input is computed as an average of $100$ samples.
One can observe the benefit of OED against a random input especially in early iterations.

\begin{figure}[tbhp!]
	\centering
	\includegraphics[width=0.9\linewidth]{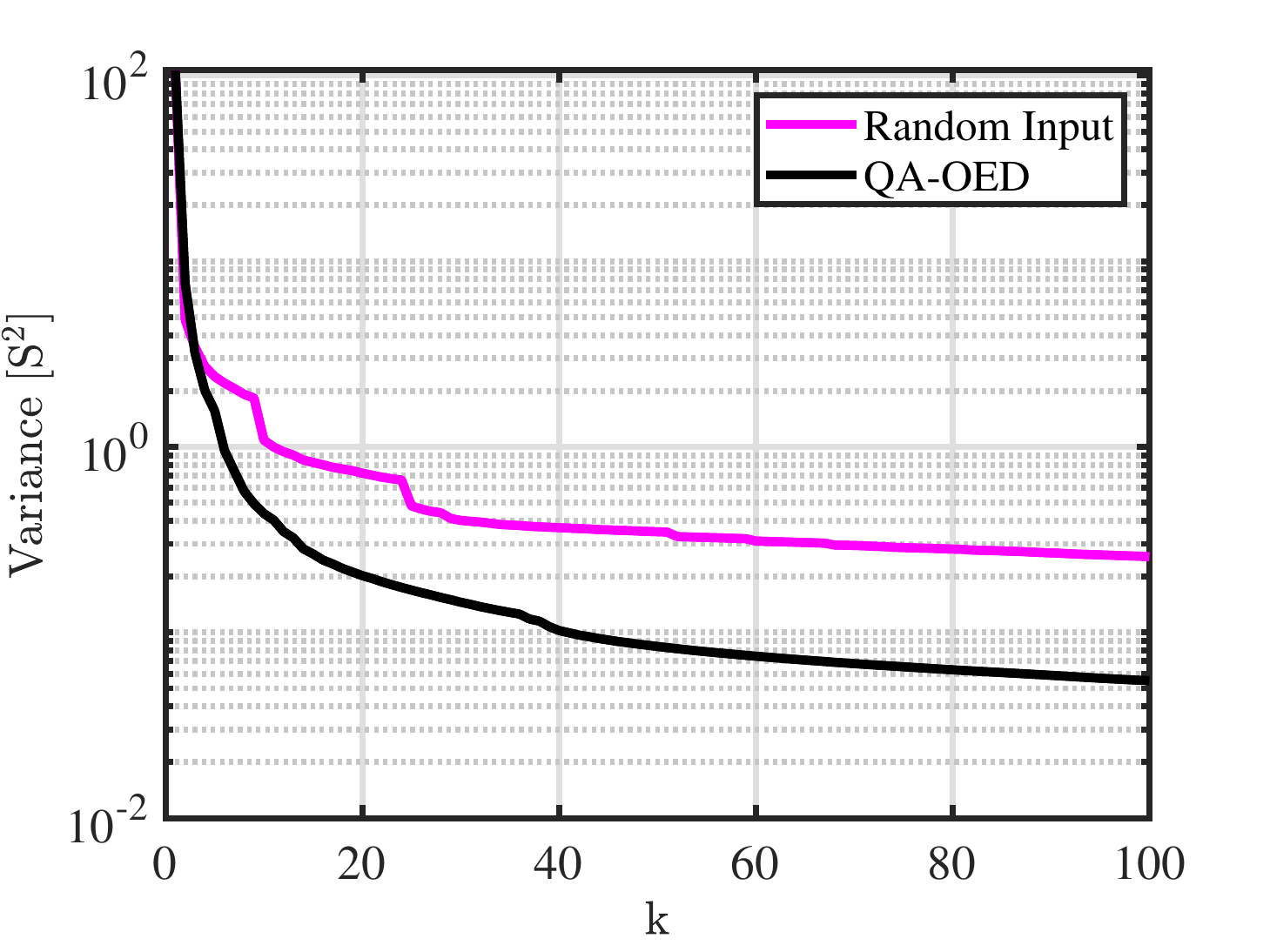}
	\caption{Expected total variance comparison between random input and QA-OED for the 14-bus power system. 
	}
	\label{fig:trace14}
\end{figure}

\autoref{fig:trace14} shows the total variance $\mathrm{Tr}(\mathcal{F}(\hat x,\hat u,\hat y)^{-1})$ with QA-OED applied to a modified 14-bus network. In order to have a larger generator-to-node ratio, we add extra generators on bus $\mathrm{9}$-${13}$. With our best approximation method.
QA-OED requires  $15.2507$ seconds for computing 200 estimation steps, while FLA-OED and ILA-OED are still not computational tractable for the $14$-bus power system. 
%{\Xu{Note that \cite{fabbiani2021identification} is a recent work on OED for power grid admittance estimation and this is the largest case that can be applied on.}}
%\Ae{Do the other methods not work for the 14 bus case? In other words, is QA OED the only method which scales to 14 buses?}{\Xu{Currently, yes, that's why I would like to explore other ways.}}

%\newpage
\section{Summary and Outlook}

This paper has presented  approximation methods for optimal experiment design operation of power grids in order to accelerate computation.
We have shown that different approximations are possible, and first numerical results indicate that a combination of inner Newton-type iterations with quadratic approximation is promising.
{However, further improving the scalability of OED seems crucial for application in practice.}
%Future work will evaluate the approximations on larger case studies and evaluate their performance on grids with a low generator-to-node ratio. 
{{Approaches based on semidefinte programming will be considered in future work.
}}

\renewcommand*{\bibfont}{\footnotesize}
	\footnotesize
\printbibliography

@IEEEtranBSTCTL{IEEEexample:BSTcontrol,
	CTLdash_repeated_names = "1"
}

@article{donti2019matrix,
  title={Matrix completion for low-observability voltage estimation},
  author={Donti, Priya L and Liu, Yajing and Schmitt, Andreas J and Bernstein, Andrey and Yang, Rui and Zhang, Yingchen},
  journal={IEEE Transactions on Smart Grid},
  volume={11},
  number={3},
  pages={2520--2530},
  year={2019},
  publisher={IEEE}
}

@article{pal2016pmu,
  title={A PMU placement scheme considering realistic costs and modern trends in relaying},
  author={Pal, Anamitra and Vullikanti, Anil Kumar S and Ravi, Sekharipuram S},
  journal={IEEE Transactions on Power Systems},
  volume={32},
  number={1},
  pages={552--561},
  year={2016},
  publisher={IEEE}
}

@article{manousakis2012taxonomy,
  title={Taxonomy of PMU placement methodologies},
  author={Manousakis, Nikolaos M and Korres, George N and Georgilakis, Pavlos S},
  journal={IEEE Transactions on power Systems},
  volume={27},
  number={2},
  pages={1070--1077},
  year={2012},
  publisher={IEEE}
}

@InProceedings{Li2010,
  Title                    = {Small test systems for power system economic studies},
  Author                   = {F. Li and R. Bo},
  Booktitle                = {IEEE PES General Meeting},
  Year                     = {2010},
  Month                    = {July},
  Pages                    = {1-4},
  Doi_disabled             = {10.1109/PES.2010.5589973},
  ISSN                     = {1932-5517}
}

@Book{Abur2004,
  title     = {Power System State Estimation: Theory and Implementation},
  publisher = {CRC Press},
  year      = {2004},
  author    = {A. Abur and A. G. Exp{\'o}sito},
  series    = {Power Engineering},
}

@Article{Hauswirth2018,
  author  = {A. {Hauswirth} and S. {Bolognani} and G. {Hug} and F. {D\"orfler}},
  title   = {Generic Existence of Unique Lagrange Multipliers in AC Optimal Power Flow},
  journal = {IEEE Control Systems Letters},
  year    = {2018},
  volume  = {2},
  number  = {4},
  pages   = {791-796},
  month   = {Oct},
}

@Article{Slutsker1996,
  author  = {I. W. {Slutsker} and S. {Mokhtari} and K. A. {Clements}},
  title   = {Real time recursive parameter estimation in energy management systems},
  journal = {IEEE Transactions on Power Systems},
  year    = {1996},
  volume  = {11},
  number  = {3},
  pages   = {1393-1399},
  month   = {Aug},
}

@Article{Bian2011,
  author  = {X. {Bian} and X. R. {Li} and H. {Chen} and D. {Gan} and J. {Qiu}},
  title   = {Joint Estimation of State and Parameter With Synchrophasors---Part II: Parameter Tracking},
  journal = {IEEE Transactions on Power Systems},
  year    = {2011},
  volume  = {26},
  number  = {3},
  pages   = {1209-1220},
  month   = {Aug},
}

@ARTICLE{Telen2013,
author = {Telen, D. and Houska, B. and Logist, F. and Van Derlinden, E. and Diehl, M. and Van Impe, J.},
title = {Optimal experiment design under process noise using Riccati differential equations},
journal = {Journal of Process Control},
year = {2013},
volume = {23},
pages = {613-629}
}

@Book{Ljung1999,
  title     = {System Identification - Theory for the User},
  publisher = {Prentice Hall},
  year      = {1999},
  author    = {L. Ljung},
  address   = {New Jersey},
  edition   = {2nd ed},
  isbn      = {978-0-136-56695-3},
}

@Article{Andersson2019,
  author  = {J. A. E. Andersson and J. Gillis and and G. Horn and J. B. Rawlings and M. Diehl},
  title   = {{CasADi}: a software framework for nonlinear optimization and optimal control},
  journal = {Mathematical Programming Computation},
  year    = {2019},
  volume  = {11},
  number  = {1},
  pages   = {1--36},
  month   = {Mar},
  day     = {01},
}

@Article{Houska2015,
  author  = {B. Houska and D. Telen and F. Logist and M. Diehl and J. F.M. Van Impe},
  title   = {An economic objective for the optimal experiment design of nonlinear dynamic processes},
  journal = {Automatica},
  year    = {2015},
  volume  = {51},
  pages   = {98 - 103},
  issn    = {0005-1098},
}

@article{fabbiani2020identification,
	title={Identification of AC Networks via Online Learning},
	author={Fabbiani, E. and Nahata, P. and De Nicolao, G. and Ferrari-Trecate, G.},
	journal={arXiv preprint arXiv:2003.06210},
	year={2020}
}

@INPROCEEDINGS{Du2020,
	author = {Du, X. and Engelmann, A. and Jiang, Y. and Faulwasser, T. and Houska, B.},
	title = {Optimal experiment design for AC power systems admittance estimation},
	booktitle = {In Proceedings of the 21st IFAC World Congress, Berlin, Germany},
	year = {2020},
}

@book{pukelsheim2006optimal,
	title={Optimal design of experiments},
	author={Pukelsheim, F.},
	year={2006},
	publisher={SIAM}
}

@article{zimmerman2010matpower,
	title={MATPOWER: Steady-state operations, planning, and analysis tools for power systems research and education},
	author={Zimmerman, Ray Daniel and Murillo-S{\'a}nchez, Carlos Edmundo and Thomas, Robert John},
	journal={IEEE Transactions on power systems},
	volume={26},
	number={1},
	pages={12--19},
	year={2010},
	publisher={IEEE}
}

@book{zhu2015optimization,
	title={Optimization of power system operation},
	author={Zhu, Jizhong},
	year={2015},
	publisher={John Wiley \& Sons}
}

@article{frank2012optimal1,
	title={Optimal power flow: a bibliographic survey I},
	author={Frank, Stephen and Steponavice, Ingrida and Rebennack, Steffen},
	journal={Energy systems},
	volume={3},
	number={3},
	pages={221--258},
	year={2012},
	publisher={Springer}
}

@article{frank2012optimal2,
	title={Optimal power flow: a bibliographic survey II},
	author={{S. Frank, I. Steponavice, and S. Rebennack,}},
	journal={Energy systems},
	volume={3},
	number={3},
	pages={259--289},
	year={2012},
	publisher={Springer}
}

@article{fabbiani2021identification,
	title={Identification of AC Distribution Networks With Recursive Least Squares and Optimal Design of Experiment},
	author={Fabbiani, Emanuele and Nahata, Pulkit and De Nicolao, Giuseppe and Ferrari-Trecate, Giancarlo},
	journal={IEEE Transactions on Control Systems Technology},
	year={2021},
	publisher={IEEE}
}

@INPROCEEDINGS{Du2021,
	author = {Du, X. and Engelmann, A. and Faulwasser, T. and Houska, B.},
	title = {Online power system parameter estimation and optimal operation},
	booktitle = {In Proceedings of the American Control Conference, New Orleans, USA},
	year = {2021},
	pages = {3126--3131},
}

@article{christie2000power,
	title={Power systems test case archive},
	author={Christie, Rich},
	journal={Electrical Engineering dept., University of Washington},
	volume={108},
	year={2000}
}

@inproceedings{lateef2019bus,
	title={Bus admittance matrix estimation using phasor measurements},
	author={Lateef, Omer and Harley, Ronald G and Habetler, Thomas G},
	booktitle={2019 IEEE Power \& Energy Society Innovative Smart Grid Technologies Conference (ISGT)},
	pages={1--5},
	year={2019},
	organization={IEEE}
}

@inproceedings{saadeh2016estimation,
	title={Estimation of the bus admittance matrix for transmission systems from synchrophasor data},
	author={Saadeh, Mahmood and Alsarray, Muthanna and McCann, Roy},
	booktitle={2016 IEEE/PES Transmission and Distribution Conference and Exposition (T\&D)},
	pages={1--5},
	year={2016},
	organization={IEEE}
}
\balance

\end{document}